# MACAO-CRIRES, a step towards high-resolution spectroscopy

J. Paufique[(1)][1], P. Biereichel[(1)], R. Donaldson[(1)], B. Delabre[(1)], E. Fedrigo[(1)], F. Franza[(1)], P. Gigan[(2)], D. Gojak[(1)], N. Hubin[(1)], M. Kasper[(1)], U. Käufl[(1)], J-L. Lizon[(1)], S. Oberti[(1)], J-F. Pirard[(1)], E. Pozna[(1)], J. Santos[(1)], S. Stroebele[(1)]

(1) European Southern Observatory (ESO), Karl Schwarzschildstr. 2, 85748 Garching-bei-München, Germany

(2) LESIA, Observatoire de Paris Meudon, 5 place Jules Janssen, 92195 Meudon, France

## ABSTRACT

High resolution spectroscopy made an important step ahead 10 years ago, leading for example to the discovery of numerous exoplanets. But the IR did not benefit from this improvement until very recently. CRIRES will provide a dramatic improvement in the 1-5 micron region in this field. Adaptive optics will allow us increasing both flux and angular resolution on its spectra. This paper describes the adaptive optics of CRIRES, its main limitations, its main components, the principle of its calibration with an overview of the methods used and the very first results obtained since it is installed in the laboratory.

**Keywords:** Adaptive Optics, Curvature, Real-Time-Computer, APD, spectrograph, VLT

## 1. INTRODUCTION

CRIRES is a first-generation instrument of the VLT; installed at the Nasmyth focus of ANTU, it will provide high-resolution spectra in the NIR (R=$10^5$ in the 1-5 micron range). Answering a need in the VLT instrumentation, it is as well the unique instrument offering such a resolution on 8-m class telescopes. High-resolution is up to now limited to the visible[1], 1-5 micron spectroscopy limited to less than $10^5$ [2] or with the limitations of FT spectrographs, and although some other experiment allow to explore such high-resolution in the infrared[3], their use is limited to longer wavelength, and to smaller telescope diameters. Therefore, CRIRES, by allowing resolving the H-lines and providing high spatial resolution, will allow studying deeply embedded stars as well as close binaries, providing new insights on the kinematics of these stars; it will as well allow unprecedented studies of solar-system objects[4,5].

CRIRES will use in normal operation a 0.2" slit size on the sky; the AO of CRIRES will provide a corrected beam for natural guide-star dimmer than $M_R$=16. The concept used for the AO of CRIRES is mostly based on a series of curvature AO-systems, developed for the VLTI and for SINFONI[6,7], SINFONI having just provided its first light[8], while the third VLTI AO is currently on its way to the mountain (first light planned for August this year).

CRIRES is to be commissioned by mid-2005. While the –cold– spectrograph is currently being integrated in the laboratory in Garching, the AO is already in test, and provided its first close-loop. In the present work we present MACAO-CRIRES design and performances and the first measurements done so far in the laboratory.

## 2. THE MACAO-AO-SYSTEM

MACAO is the acronym of Multi-Applications Curvature Adaptive Optics. MACAO-CRIRES is based on a 60-actuators deformable mirror (DM); the optics can compensate at a 3σ level the atmospheric wavefront curvature averaged over any actuator area, up to a 1" seeing[2]. This could ideally provide a Strehl ratio in K band above 55% for 1" seeing (with an infinite bandwidth correction), as can be seen on figure 1.

---

[1] Contact: jpaufiqu@eso.org
[2] All seeing and coherence time ($\tau_0$) values are given at 500 nm in this paper; Strehl are evaluated in K-band, unless specified.

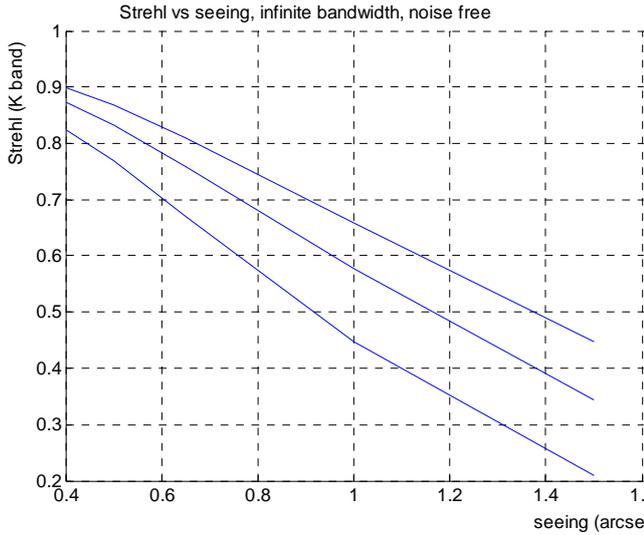
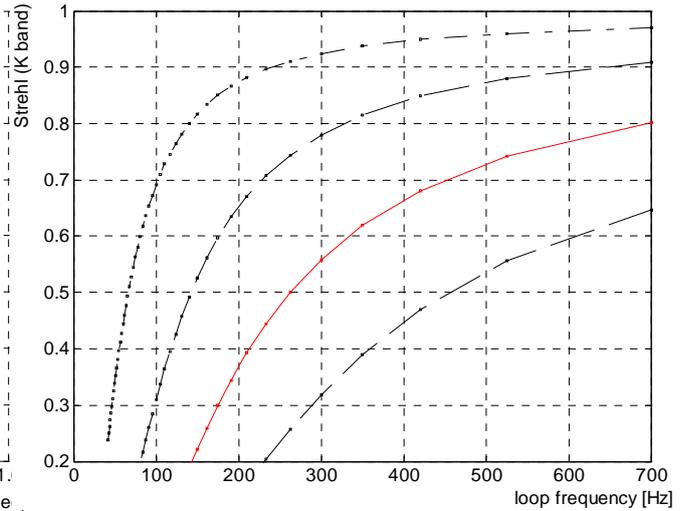

Figure 2: Strehl ratio vs. seeing, for three different numbers of Zernike modes corrected (40, 60, 80); bandwidth is supposed infinite, DM stroke and noise are not limiting factors.

Figure 2: Strehl loss vs. loop frequency, for different $\tau_0$. All Zernike modes are corrected, only delays (0.5 ms) and integration time are considered; curves are indicative only for low Strehl values. From bottom to top, $\tau_0$=2, 3, 5 and 10 ms.

The wavefront sensor integrates the curvature signal during 1 loop cycle of duration $D_{loop}$, and the signal is then processed by the Real time computer (RTC) during $D_{RTC}$, before being applied by the DM high voltage electronics, which lasts $D_{HVA}$. This introduces a delay $\Delta$ in the correction which is:

$$\Delta = D_{loop} + D_{RTC} + D_{HVA}$$

The CPU and electronics delays were specified to be altogether less than 500 µs. Actual devices provide respectively 350 µs and 150 µs, for a sum which meets the specification. Taking these parameters into account and neglecting the dynamical response of the DM, we may estimate the loss of Strehl ratio due to the refresh frequency only, as seen in figure 2.

Of course, photon noise plays as well a major role in our systems, and imposes to adapt the loop frequency (or the loop gain which is equivalent in our system), depending on the guide star magnitude. The expected optical throughput of MACAO-CRIRES at the Nasmyth focus of one of the VLT is around 40%, or about 730 photo-counts /ms for an $m_V$=10 G-star on each sub-aperture.

A realistic estimation of the performance considers spatial fitting error, delays (integration of the signal, calculation and application of the correction), optical aberrations ("high" spatial frequencies and non-common path wavefront errors) and sensor noise. Such features were implemented in a simulation package developed at ESO in the early stages of the MACAO systems design studies[9]. The loop frequency was optimised at each flux, so as to maximise the Strehl ratio. The corresponding performances are drawn on figure 3. The Strehl ratio can be maintained above 50% for stars down to magnitude 14 for good seeing conditions (0.65", $\tau_0$=4 ms).

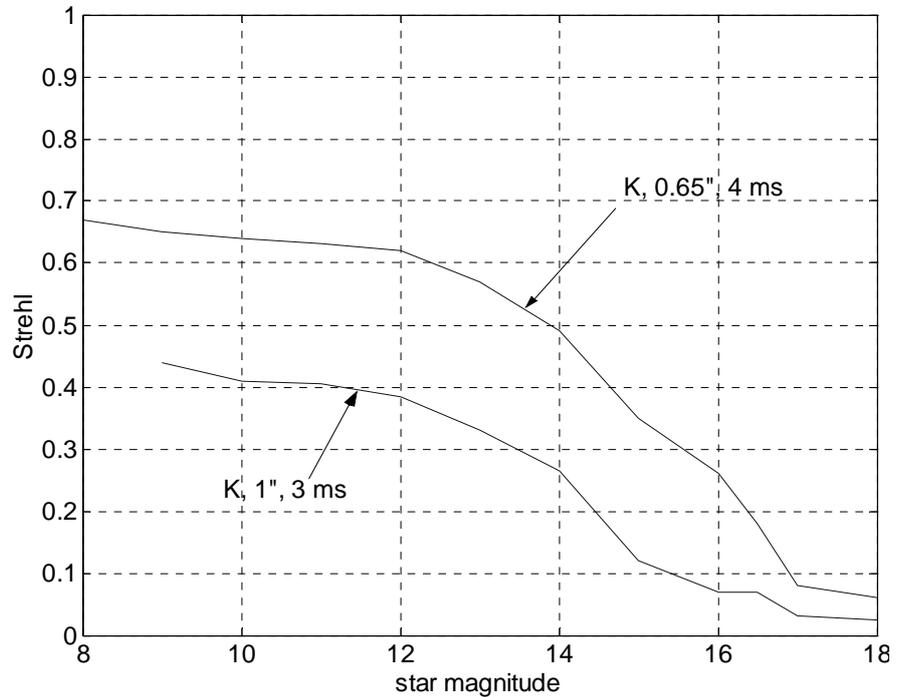

Figure 3: expected K-band performance of the MACAO-SINFONI/CRIRES systems. Upper line: seeing=0.65", $\tau_0$=4 ms, bottom line: seeing=1", $\tau_0$=3 ms

## 2.1. The CRIRES case

In CRIRES, the slit size being 0.2", a relevant parameter is the encircled energy within 0.2". We made some simulations, by modelling the PSF with a diffraction limited spot with an intensity multiplied by the Strehl ratio, and adding to this the rest of the energy modelled by a Gaussian seeing disk. The incoming fluxes were scaled from the 0.65" case, which leads to some inaccuracy for very good seeings (this is visible in figure 4 especially in M-band). For 2003's usual seeings (0.5" to 1"), the gain is weak in J-band, and of the order of 2 for the K- to M-bands.

Then, the performance of the system has been evaluated through the comparison between the corrected and non-corrected flux going through the slit. This estimation is conservative, as we did not take into account the reduction of the useful size of the source along the slit axis, which benefit in terms of sky background to the AO images. Despite this pessimistic estimation, an improvement of around one magnitude is observed.
This performance is strongly related to the accuracy of the alignments and to the quality of some critical components, so that special care has been given to all the optical components of the system, to keep low the corresponding losses.

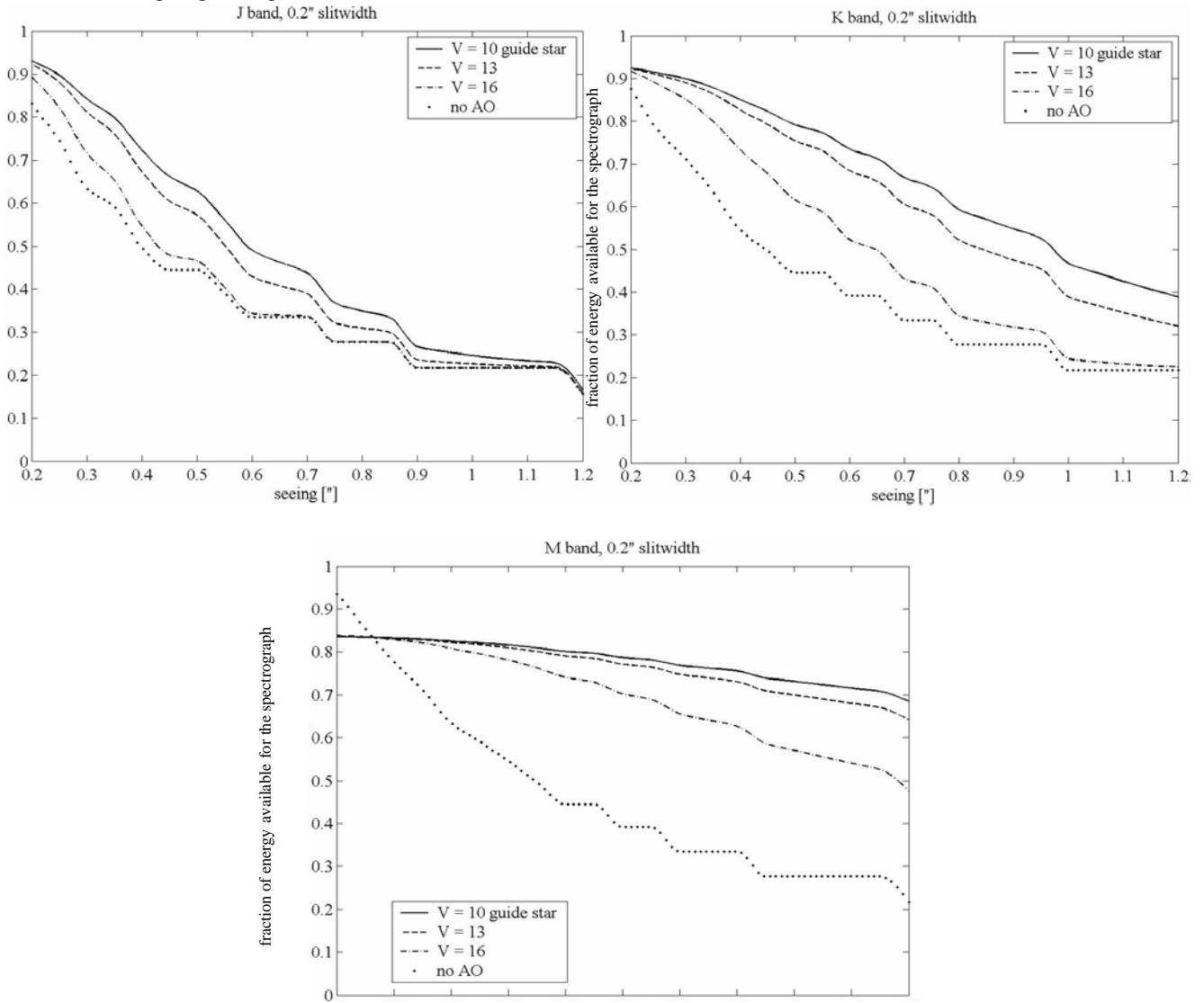

Figure 4: compared performances between corrected and uncorrected images. These simulations do not take into account the gain due to the narrowing of stellar images along the slit.
For extended objects, the gain is very dependent on the target; the contrast between features being increased, the sensitivity of the instrument will be improved, but no general comparison can be made.

### 2.2. Overview of the system

A calibration unit (for spectrographic and adaptive optics purpose) is located at the Nasmyth focus, before a 3-mirrors derotator. It includes several calibration sources and absorption cells, as well as artificial stars, allowing turbulence-free calibrations of the AO system. The whole warm optics assembly is located on a breadboard, which can be covered by dark panels, to allow day-time calibrations.

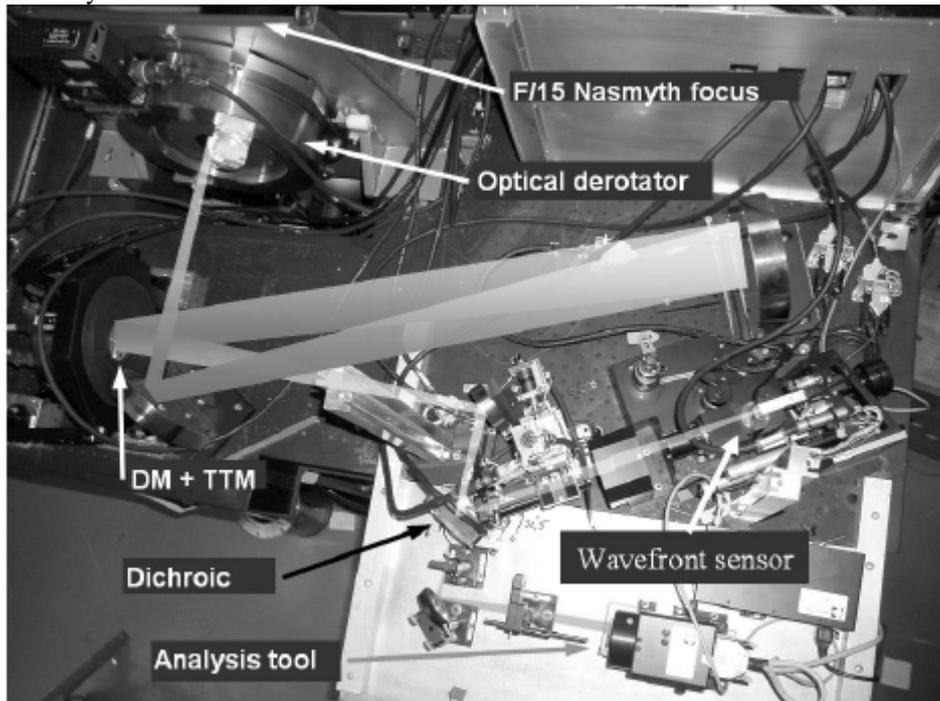

Figure 5: top view of the warm optics of CRIRES (including the wavefront sensor); the relay optics is located after the 1 arcmin field derotator. The picture is approximately 1.5 m wide.

#### 2.2.1. The relay optics
A spherical mirror tilted by 3.2º transfers the Nasmyth focus with a 1:1 magnification, its astigmatism being corrected by two other long radius spherical mirrors. The compensation of the astigmatism is achieved in the centre of the field, while some field aberrations still remain.
The relay embeds the DM, which is located in a 60 mm diameter pupil plane; the incidence angle is 8º, generating a slightly elongated footprint on the DM (1%).

#### 2.2.2. The deformable mirror
The DM is based on a bimorph technology and developed by CILAS, for all MACAO applications, declined in 2 flavours[5]. CRIRES and SINFONI[6] use the same 110 mm mechanical diameter version of DM. The mirror has two main resonant frequencies, at about 230Hz and 880 Hz, corresponding respectively to a trefoil and focus modes.

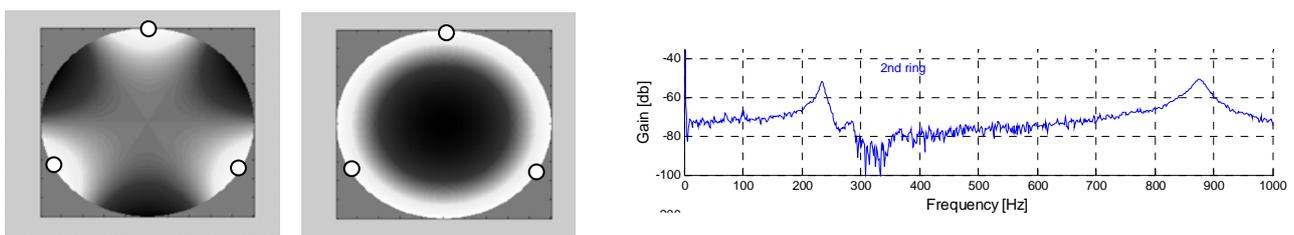

Figure 6: resonant modes of the mirror (synthetic images; the support points are represented by small circles). The associated resonant peaks are visible on the transfer function (right).

The trefoil is caused by the 3-points support: three V-grooves pins sandwich the edge of the mirror, being fixed points for the mirror. The focus is the natural vibration mode of the free mirror. These modes are superposed to a piston mode (as the support points are fixed and not the centre), in which we have no interest, this piston mode producing no aberration to the images of CRIRES.

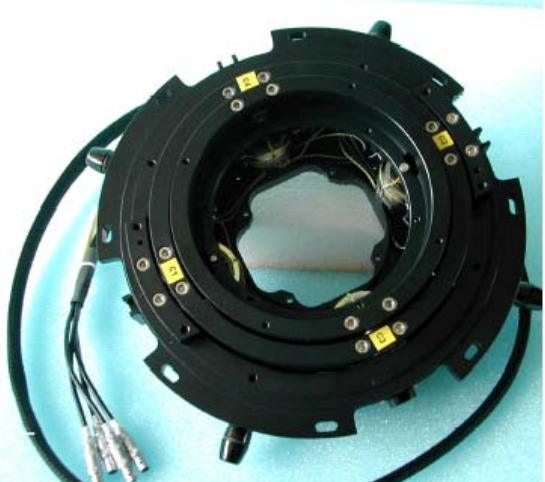

Figure 7: tip-tilt mount, front view; the four linear actuators are visible inside the mount, at 45º from the handles.

The mirror is mounted on a gimball tip-tilt mount, so that the tip-tilt component of the correction applied to the mirror can be offloaded. The gimball mount was developed by the LESIA, and shows a bandwidth intentionally limited to 100 Hz, allowing correcting the whole tip-tilt with the mount itself, though in all our systems, an even lower operating bandwidth was chosen, in order to avoid any transfer of energy to the resonant modes of the mirror (typically 10 times slower than the resonant modes of the mirror) and minimise interferences between the two loops (see Control loop section).

### 2.2.3. The dichroic

Located only 5 cm ahead of the focus, the dichroic is based on the NAOS dichroic design and provides a transmission of the IR part of the beam close to 90% (with 5% of variation) over the 0.95-5.3 micron range, and a reflection of the visible light (450 nm-900 nm) of 96%, excepted close to the cut-off wavelength.
At the visible focus, the pupil is then located at a finite distance, with an f/14.8 speed. The MACAO system having been designed to work with an f/47 telecentric beam, the focus must be re-imaged with a 3.2 enlargement to fit this requirement. The designed relay optics includes as well an off-axis functionality: it allows the AO to track on a star up to 30 arcsec off the pointing axis of the telescope (e.g. the observed object).

### 2.2.4. The re-imaging optics

A first telecentricity lens re-images the pupil at infinity. This allows the scanning lens located right after, to have the same incidence angle, whatever is the position of the star in the field.
The scanning lens has four functions:
∴ To collimate the beam from the visible focus,
∴ to set the distance between the re-imaged pupil and the collimator at $f_{collimator}$, the focal distance of the imaging lens,
∴ to pick the guide star within the field of view, through a 3-axis translation stage,
∴ and to feed the AO with tip-tilt low- frequency offsets, coming from a slit viewer, designed to compensate for differential refraction and spectrograph flexures.

The telecentric imaging lens focuses the light at the membrane level. The membrane is oscillating at its resonant frequency, i.e. 2100 Hz, excited by a speaker monitored by the RTC.
When the membrane is flat, the pupil image stays at infinity. The wavefront sensor will focus it on the lenslet array.

### 2.2.5. The wavefront sensor

A spherical mirror collimates the light coming from the membrane mirror, while re-imaging the pupil at the entrance of a beam expander. The beam expander is based on a two off-axis parabola afocal lens. It enlarges the beam to a diameter of 14 mm, while forming a real image of the pupil at the level of a lenslet array[10]:

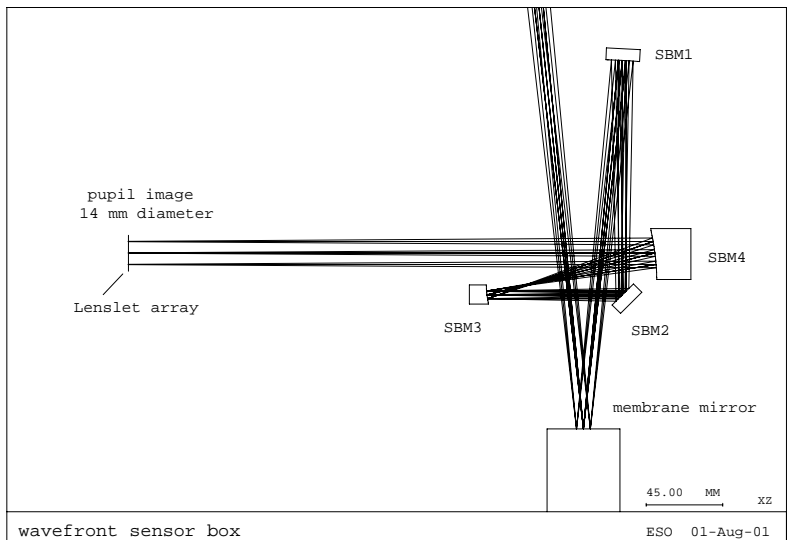

Figure 8: optical design of the MACAO curvature wavefront sensor. The beam being telecentric on the membrane, tilted beams have the same incidence angle on the membrane (three different tilt positions are drawn here). The membrane is flat in this design.

After the membrane mirror, aberration in the optics will essentially lead to a blurring of the pupil images, without affecting the measurement of curvature itself. Therefore, the optical quality of the beam expander has been relaxed up to 120 nm rms, without affecting the performances.

### 2.2.6. The lenslet array

The lenslet array is the point at which the light is split in 60 distinct optical channels, to be processed later on.

We chose a two-step assembly: a first lenslet reproducing the geometry of the DM and focusing the light of each subaperture on a second lenslet, acting as a Fabry lens which images the subaperture on a fibre entrance. This Fabry lens prevents the system suffering from injection losses related to the tip-tilt at the level of the subaperture.

For the first lenslets, Heptagon (Swiss) improved their technological performances to manufacture high depth lenslets (up to 25 micron), the laser writing of a master before replicas are produced out of it[9]. The second lenslet array has been realised by Microfab, an american company manufacturing lenslets through a process of micro-deposition of droplets (inkjet-like), producing high quality of geometrical properties of the pattern (in our case, half-ball lenses, 0.7mm of diameter, about 0.5 mm of useful aperture). The light is then injected in 60 fibres, positioned with a high accuracy in a frame at the focus of the ball lenses. The bundle of 60 fibres guides the light towards the sensors.

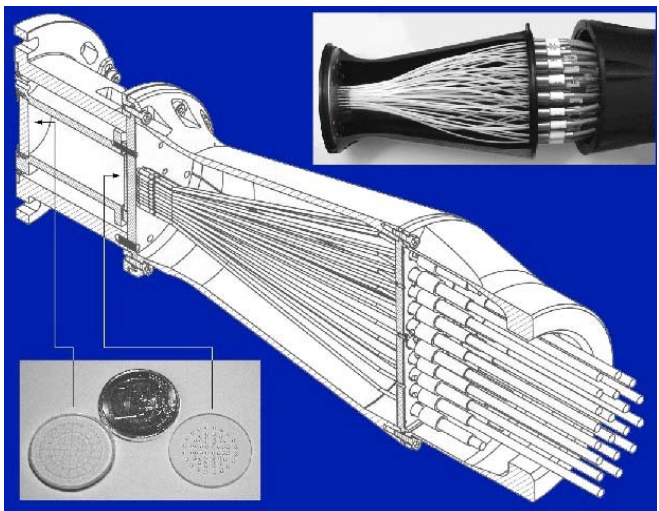

Figure 9: front-end assembly of the fibre bundle; the sixty fibres are assembled in a very narrow arrangement, to fit the geometry of the vertexes of the 60 lenslets. At the other end, each fibre has an FC connector.

### 2.2.7. The APD cabinet

The light is then directed towards 60 individual photon counters (Avalanche PhotoDiodes, APD) from Perkin-Elmer. The 60 APD are arranged in a rather compact configuration (the rack is 600x600x800 mm wide).

Despite this density, the APDs are all rather easily exchangeable.

By the end of 2005, 360 APDs will be in operation in Paranal, with an expected failure rate of 8-20 units per year. Some of these failures are predictable, allowing a preventive maintenance, some are not. Although one failing APD does not prevent to use the system with degraded performances (allowing for example avoiding in some cases to interrupt night operation), it remains important being able to exchange a unit within a short time.

The signals of the APDs are relayed by an electronics located inside the rack directly to the RTC's VME bus.

### 2.2.8. The RTC

The oscillating membrane produces a signal which modulation is proportional to the local wavefront curvature. This signal, collected by avalanche photodiodes (APD), is sent to the RTC, which processes the signals with the pseudo-inverse of an interaction matrix calibrated with a procedure which has been refined during the development of the MACAO-VLTI and MACAO-SINFONI systems.

The RTC is based on a PowerPC (PPC), connected to the High voltage amplifier (HVA) and the tip-tilt stage through a serial I/O fast data fibre link.

The control loop task may take several modes: the open loop mode, in which the RTC can send data to the DM and motors, read the data coming from the WFS; the feedback mode, in which the system closes the loop, exchanges data with the command server (offload of tip-tilt, movement of motors, …), a calibration mode, in which interaction matrices may be computed (commands sent to the mirror and values read from the WFS) and a test mode, in which the RTC offers some maintenance tools to the user (generation of sine or square excitations of the mirror, …).

The command server task of the LCU is responsible for exchanging commands and data with the RTC over the VME bus and with the Workstation over the network. It waits for a Central control software (CCS) message and then it takes appropriate actions which, normally, will cause one or more commands to be sent to the RTC for synchronously execution, and then returns results and errors through the CCS messaging system. It spawns the Loop Monitoring task, the Dataflow server task and initializes all the libraries. This software includes part of the standard VLT software distribution.

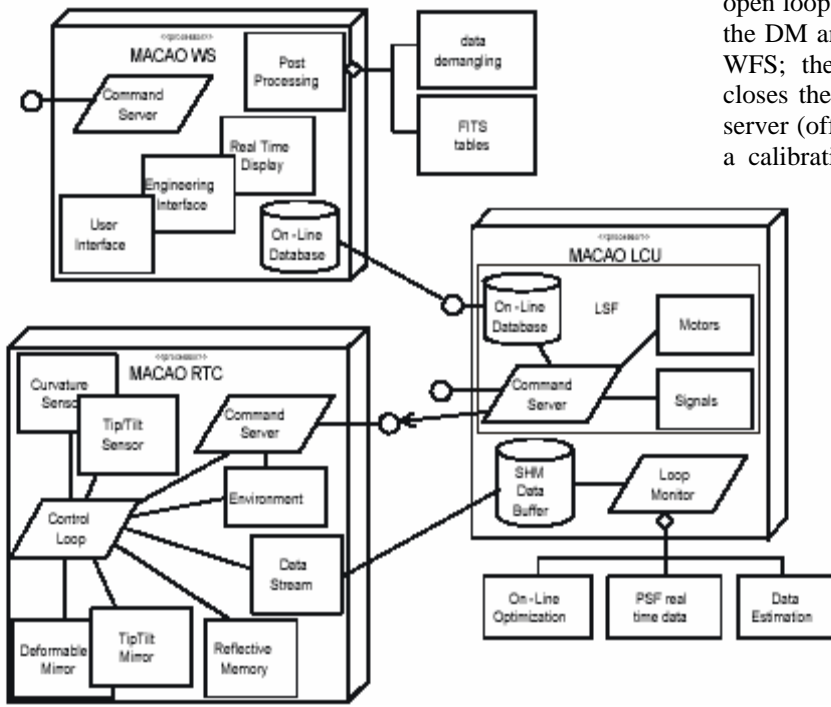

Figure 10: MACAO RTC software architecture; the software is divided in three blocks, each running on a different CPU.
The connection between MACAO-RTC and MACAO-LCU is VME based. The connection between MACAO-WS and MACAO-LCU is TCP/IP based.

The Workstation software is implemented using the VLT standard software. It exposes a programmatic interface (the command server) and a UI interface for maintenance and testing. Final data post-processing takes place in the Workstation.

### 2.2.9. The HVA and the TTM

The RTC is sending values to the HVA electronics through a direct fibre link. The 62 numbers sent at each refresh are converted into 62 voltages to apply simultaneously. The DM high voltages are produced by an electronics developed by 4DE, Germany; 4 boards with 15 outputs each are plugged in a rack linked to the DM. The HVA should be able to stand a 1 kHz refresh; driver and compatibility problems limit currently the actual refresh to 700 Hz in the case of CRIRES.

The command to the TTM is sent by the RTC, and an internal loop is driven by electronics specifically developed by the LESIA for the TTM.

### 2.3. The MACAO-CRIRES control loop

The membrane mirror oscillation frequency being too high for the RTC to handle it, the RTC accumulates several frames to reduce the frequency. The loop frequency is then obtained by accumulating 6, 5, 4 or 3 frames, resulting in loop frequency of 350Hz, 420Hz, 525Hz or 700Hz. The diagram in figure 13 represents the control loop, running at one of the above cited frequencies.

The Curvature Wavefront Sensor block is responsible of accumulating the frames to match the loop frequency and to compute the curvature signal by means of the following formula:

$$c(n) = \frac{A(n) - B(n)}{A(n) - B(n) - 2 \cdot sky(n)}$$

where 'n' is the sensor channel (from 1 to 60), 'A' and 'B' are the counts on the two half cycles and 'sky(n)' is the sky background measured on the channel 'n'. The object stores into real-time buffers each intra- and extra-focal count at the loop frequency.

After that it is possible to inject a disturbance in curvature space, to simulate the APDs in the laboratory or to add artificial disturbance to an otherwise undisturbed signal, e.g. a static curvature pattern may be used to compensate for non-common path aberrations (especially astigmatism in the case of CRIRES). However, the linear range of the sensor is very limited and the system can account only for very small closed-loop offsets. At this point, the resulting curvature is stored into real-time buffers for analysis and diagnostics.

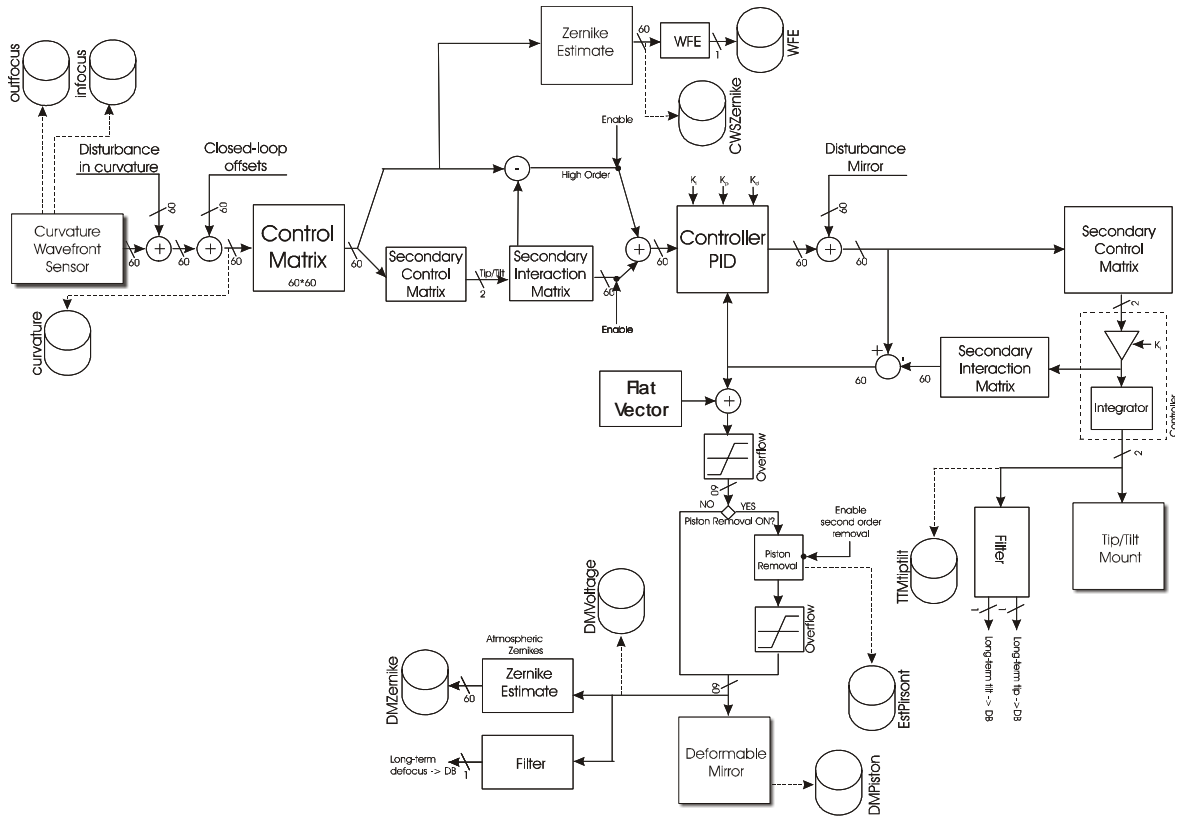

Figure 11: architecture of the MACAO control loop

The control matrix[3] is then used to project from sensor space to mirror space, i.e. to reconstruct the wavefront. The system is able to manage tip/tilt control separately from the high-order modes: it is thus possible to close the loop only in tip/tilt mode, only in high-order mode or both[4]. By means of the secondary control matrix[5] the RTC projects the residual voltages to extract the tip/tilt information and separates tip and tilt from the rest. Depending on which mode is selected, the PID controller receives either a complete residual vector, or a tip/tilt only vector or a high-order vector.

---

[3] The control matrix is the truncated pseudo-inverse of the interaction matrix, usually filtering piston mode only.

[4] This is a feature not used in CRIRES (used for Laser Guide Star mode or a extern tip/tilt sensor mode).

[5] The secondary interaction matrix is the map between the DM voltages that produces tip and tilt and the corresponding position of the TTM.

The upper part of the diagram shows the Zernike estimation: this allows the RTC to estimate the achieved correction by computing the wavefront error. Residual Zernike coefficients are also stored into the real-time buffer for additional statistics.

The controller is a simple PID, identical for all the 60 channels. The proportional and derivative part did not show significant improvements, thus only the integral part is normally used.

After the controller it is possible to add another disturbance in mirror space; we use this to study the behavior of the DM by replaying a known sequence of mirror commands, or for instance, to reduce hysteresis effects while measuring the system influence functions.

The result is the next configuration of the mirror. What follows is mainly the management of the tip/tilt offload: the DM has limited stroke and tip/tilt is a mode that demands a significant stroke, almost entirely located on the outer ring electrodes. By offloading part of all the tip/tilt corrected by the DM to the TTM, one can increase the dynamics of the system. The command vector is projected on the tip-tilt modes as delivered by the TTM, and the projection is split in two parts: one is sent to the TTM, the other is sent to the DM.

All these computations have been carried out with a mirror flat surface as reference: the vector that produces such surface is then added and a check against electrode saturation is carried out: if too many electrodes (by default 10) stay in saturation for a too long time (2 seconds) the RTC aborts and opens the loop. The final result is applied to the mirror and stored as well into real-time buffer.

Few other functions complete the control loop: the RTC decomposes the voltage pattern in Zernikes to enable reconstruction of the atmospheric parameters. Running average filters are used to compute the long term tip/tilt of the tip tilt mount and defocus of the DM to be potentially offloaded to the telescope.

### 2.4. Calibration

All the calibrations of the system are done through an engineering interface, using mostly script commands sent to the RTC and the LCU from the Workstation. The interface opens on a starting panel (see figure 14) which allows to access all calibration and control panels directly from the menu bar. From this interface, we monitor several AO-key parameters (curvature, flux, voltages applied, …) as snapshots or averaged values, and all RTC calibration panels are accessible from its menu bar. It is a powerful tool developed to calibrate and maintain the MACAO systems, and therefore benefits of almost two years of continuous improvement through the users' feedback and the development of the three different projects using MACAO systems.

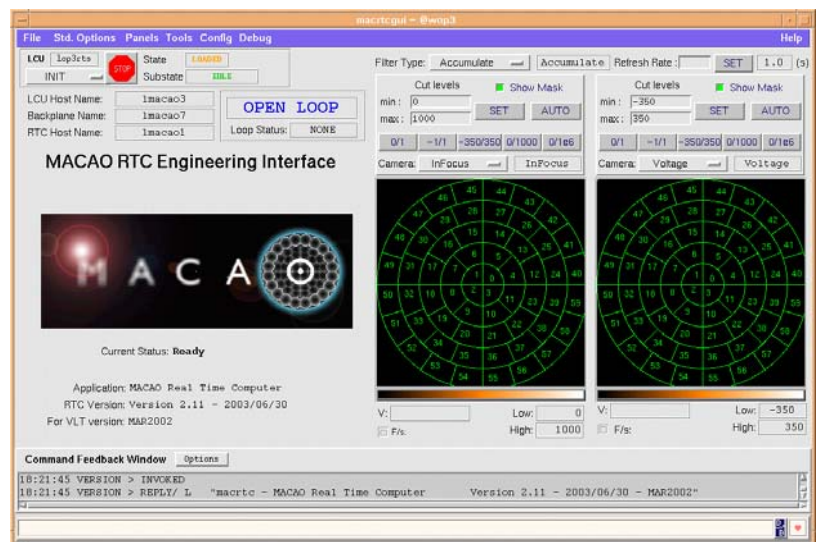

Figure 12: starting panel of the RTC engineering interface

Once the system is operational, the basic calibration of the system consists in the following steps:
- ∴ Membrane phase lag tuning; this has as object to tune the phase delay between the speaker exciting the membrane and the curvature signal provided by the APDs. This is done by minimising the curvature measured (90º of lag between excitation and measure), and by removing 90º to the value obtained.
- ∴ Pupil alignment; this is done by scanning a range of gimball positions, while recording the flux at the level of the outer electrodes, the best position being determined by balancing the flux of the outer subapertures.
- ∴ Radius of curvature (RoC) calibration; this is done by applying a known tilt to the beam, and measuring the amplitude voltage applied to the membrane speaker for which the outer electrodes see a sharp transition of flux (the tip-tilt induces a shift of the pupil proportional to the tip-tilt and inversely proportional to the RoC). A scaling law proportional to the inverse of the voltage is applied to determine the RoC for a given excitation voltage. An alternative way is to measure the curvature signal provided by the axial displacement of the scanning lens. This signal is directly related to the radius of curvature of the membrane mirror.
∴ Interaction matrix measurement (see the following section)
∴ Flat vector measurement.

In a few words, an interaction matrix is produced by recording the 60 response vectors of the wavefront sensor to a successive excitation of the 60 individual electrodes of the mirror. The result allows to identify the eigenmodes of the system, using the svd method; reproducing these modes, we may measure them more accurately, especially the modes which are badly sensed by the wavefront sensor. It is especially important to calibrate these modes properly, because a weak signal produced on the curvature sensor along these modes implies already a high voltage to be applied on the DM to compensate for it: in close loop, the miscalibration of these modes may lead to an unacceptable increase of the power used for the required correction -without any improvement in the quality of the wavefront correction-. Exciting the mirror specifically along these poorly sensed modes allows decreasing the calibration error which is made on these modes. The method is described in detail in[11,12].

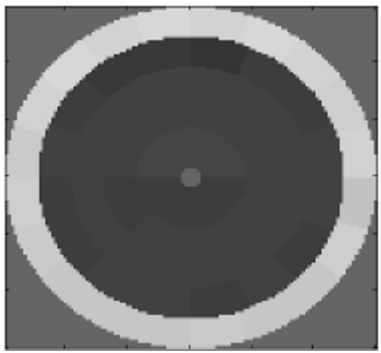

Figure 14: voltage pattern for a piston term. Outer electrodes are submitted to the highest voltages

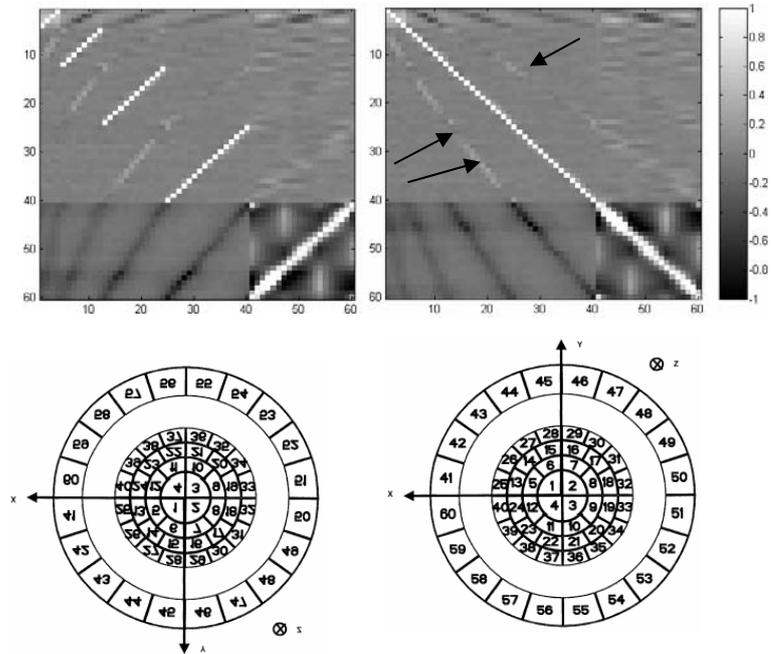

A filter is as well applied, by removing the lowest eigenvalue mode of the system from the inversion of the interaction matrix. This so-called "piston mode" is producing the lowest curvature possible on the wavefront sensor, and corresponds to a top hat-like shape of the mirror voltages, as illustrated in the figure 15.

Figure 13, left: a raw interaction matrix and the image of the pupil on the lenslets (mirrored), right: the same IM numerically mirrored, and the corresponding virtually restored pupil image on the lenslets. On each ring, lenslets and DM have a reversed ordering.
The X-axis is the DM actuator axis, the Y-axis is the curvature signal axis. The last twenty lines display the signal produced at the edge of the pupil (mostly local tip-tilt); on this matrix, a slight shift and rotation of the DM-lenslet pupil matching is noticeable (about 10% of a subaperture), by noticing the non-uniformity of the close neighbors of the diagonal sub-pupils (arrows).

This mode being not sensed by the wavefront sensor, it is of prime importance to filter it before applying any command to the mirror. This is done by zeroing numerically the lowest eigenvalue of the svd-decomposed interaction matrix, before inverting this matrix to get the control matrix (pseudo-inversion).

With respect to the MACAO-SINFONI system, the optical beam of CRIRES has one reflection less between the DM and the lenslet array, producing therefore a mirrored image of the pupil. The RTC must therefore reorder the signal coming from the APDs, in order to present the same pattern while acquiring interaction matrix, or displaying curvature on the user interface. This is shown on figure 16.

**2.5. Open loop operation**

CRIRES is as well designed to perform open loop operation, when an IR target is observed, without optical source close enough to help for AO-operation. In this case, MACAO is able to keep the mirror flat for one hour or more without degrading significantly the optical performance of the system.

## 2.6. Test bench

A turbulence generator has been installed to introduce a controlled turbulence in the system, and to give a performance estimation as close to the final performance of the system as possible. The loop could be close recently, with a Strehl ratio of 35 % for 0.65" of seeing with a wind speed approximately corresponding to $\tau_0$=3 ms. Several calibrations and optimisations remain to do, for which the experience of SINFONI is a noticeable help.

Due to the scheduling of our infrared test camera (ITC), the measure of the output wavefront has initially been done in a re-imaged pupil plane, with a high resolution wavefront sensor (HASO 64, 64x64 Shack-Hartmann sensor from Imagine-Optics), allowing to take snapshots of the output wavefront. 60x60 pupils were illuminated, which gives approximately 2900 subapertures for the measurement, 40 times the number of subapertures of the MACAO sensors. Nevertheless, due to some high spatial frequencies present on the phase screens (about 10 mm$^{-1}$), it was not possible to get an accurate measurement of the wavefront over one full rotation of the phase screens (i.e. 10 s). Even with this drawback, the characterisation of the open-loop and close-loop wavefront was possible, allowing assessing the performance of the system. For a bright star ($m_V$=10.5), 0.65" seeing, a $\tau_0$=3 ms (estimation), and a RoC of 50 cm, the correction quality reached 35%. The system was not yet fully optimised, especially the RoC used. The full optimisation of the MACAO-VLTI systems provides Strehl ratio close to 55% in the same conditions, which shows the progress we are able to do in the final performance of the system.

## 3. CONCLUSION

The MACAO-CRIRES system has shown its capacity to correct a turbulent wavefront, providing diffraction-limited images to the IR-focus, improving the sensitivity of the instrument by more than a factor two. The next months will be dedicated to the full verification of the system parameters (field of view of the wavefront sensor, ability to correct the field aberration with close-loop offsets, determination of the scanning lens-membrane mirror coordination, …), and to the optimisation of the system, before the AO delivers its corrected beam to the spectrometer.

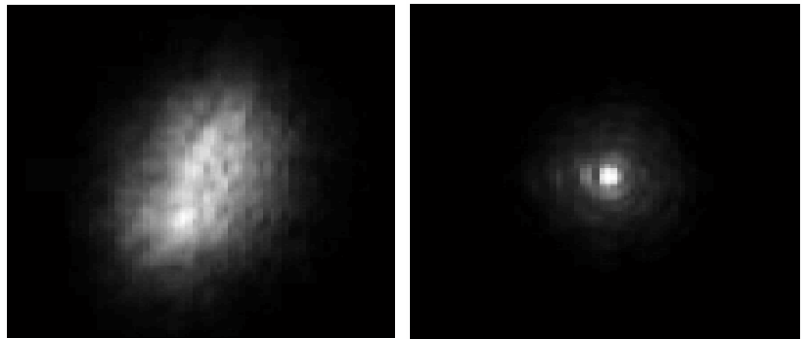

Figure 15, left: open loop, right: first closed loop; reconstructed PSF from wavefront measurements. The short integration time was not enough to circularise the open loop PSF.

## 4. ACKNOWLEDGEMENTS

The authors would like to acknowledge the teams of MACAO-SINFONI and MACAO-VLTI for all the work done on the MACAO systems from which CRIRES benefits now, with a special mention for R. Arsenault for his pertinent comments on this paper.

1 H. Dekker et al., 2000, "Design, construction, and performance of UVES, the echelle spectrograph for the UT2 Kueyen Telescope at the ESO Paranal Observatory", Proceedings SPIE [4008-534]
2 Hinkle, Kenneth H. et al., "The Phoenix Spectrograph at Gemini South", Proceedings SPIE [4834-353]
3 J. H. Lacy et al., "TEXES: A Sensitive High-Resolution Grating Spectrograph for the Mid-Infrared"
PASP, 114:153–168, 2002 February
4 H. Käufl, et al., "CRIRES: a high resolution infrared spectrograph for ESO's VLT", Proceedings SPIE [5492-#60]
5 R.J.Dorn et al., "The CRIRES InSb megapixel focal plane array detector mosaic", Proceedings SPIE [5499-#57]
6 R. Arsenault, et al., "MACAO-VLTI adaptiv optics systems performance", Proceedings SPIE [5490-#06]
7 H. Bonnet, et al., "Implementation of MACAO for SINFONI at the VLT, in NGS and LGS modes", Proceedings SPIE [4839-#39]

8 H. M. Bonnet et al., "MACAO for SINFONI: performance optimization and characterization", Proceedings SPIE [5490-#99]
9 T.V. Craven-Bartle et al., "Computer simulation comparison of CCDs and APDs for curvature wavefront sensing", Proceedings SPIE [4007-444]
10 J. Paufique et al., "Individual testing of replicated micro-lenses: wavefront error and psf estimation." Proceedings SPIE [5490-#131]
11 S. Oberti et al., "Calibration procedures for a curvature sensor/bimorph mirror AO system: lessons learned from MACAO and SINFONI", Proceedings SPIE [5490-#100]
12 M. Kasper et al., "Fast calibration of high-order adaptive optics systems", JOSA A, Volume 21, Issue 6, 1004-1008 June 2004